\documentclass[aps,prl,preprint,groupedaddress]{revtex4}

\begin{document}


\title{Gravitational waves in the presence of a dynamical four-form}


\author{Patrick Das Gupta}
\email[]{patrick@srb.org.in}
\affiliation{Department of Physics and Astrophysics, University of Delhi, Delhi - 110 007 (India)}


\date{\today}

\begin{abstract}
Propagation of weak gravitational waves, when a dynamical four-form is around,  has been investigated. Exact, self-consistent solutions corresponding 
 to plane, monochromatic gravitational waves have been studied.
\end{abstract}

\pacs{}

\maketitle
\section{Introduction}
There are strong observational evidence in support of an accelerating universe.  A small but positive cosmological constant or some dark
 energy could be responsible for the accelerated expansion [1,2 and references therein]. Is there a geometrical origin to the dark energy? 
I\lowercase{N SPECIAL RELATIVITY, $\eta_{\mu \nu}$ AND $\epsilon_{\mu \nu \rho \sigma}$ ARE BOTH INVARIANT  UNDER PROPER} L\lowercase{ORENTZ TRANSFORMATIONS.}
 S\lowercase{PACE-TIME GEOMETRY  REPRESENTED BY  $g_{\mu \nu}$ (THAT BECOMES $\eta_{\mu \nu}$ IN A LOCAL INERTIAL FRAME) IS DYNAMICAL} in general relativity (GR).
 What about
$\epsilon_{\mu \nu \rho \sigma}$ in GR?
Taking this as a cue, we extend GR by considering a dynamical four-form $\tilde {w}$  [3],        
 $$\tilde {w} = {1\over {4!}} w_{\mu \nu \rho \sigma}  \tilde{d} x^\mu \wedge   \tilde{d} x^\nu \wedge   \tilde{d} x^\rho \wedge  \tilde{d} x^\sigma \eqno(1a)$$       
 so that,
 $$w_{\mu \nu \rho \sigma} (x^\alpha)= \phi (x^\alpha) \epsilon_{\mu \nu \rho \sigma}\eqno(1b)$$
 with $\phi (x^\alpha) $ \lowercase{BEING A SCALAR-DENSITY OF WEIGHT +1, THAT TRANSFORMS AS}   $\phi \rightarrow \phi /J $, \lowercase{UNDER A GENERAL COORDINATE TRANSFORMATION.}
       
Dynamical scalar-densities have been considered previously in the context of non-metric volume-forms [4,5 and references therein]. Our approach however is different, stressing on its  
 association with Chern-Simons extension and dark energy [3,6]. 
W\lowercase{E STUDY THE  ACTION} $S$, 
$$S= - {{m^2_{Pl}}\over{16 \pi}} \int {R \sqrt {-g} d^4x}+ \int {L \sqrt {-g} d^4 x}\  + $$
$$\ \ \ \ \ \ + {{A}\over{4!}} \int {\phi \ w^{\mu \nu \alpha \beta}_{\ \ ;\lambda}  w_{\mu \nu \alpha \beta}\ ^{;\lambda} \ d^4 x }
+ B \int {\phi (x) d^4 x} + S_{GCS} \eqno(2)$$
with a Chern-Simons  part $S_{GCS}$, inspired by Jackiw and Pi's paper [6],
$$S_{GCS}= H\int {w^{\mu \nu \alpha \beta} [\Gamma^\sigma_{\nu \tau} \partial _\alpha \Gamma^\tau_{\beta \sigma} + {2\over{3}} \Gamma^\sigma_{\nu \tau}\Gamma^\tau_{\alpha \eta}\Gamma^\eta_{\beta \sigma}]\phi_{;\mu} \ d^4 x} \eqno(3)$$    
where $L$ \lowercase{IS THE} L\lowercase{AGRANGIAN DENSITY FOR MATTER FIELDS} while  
 $A$ and $B$ \lowercase{ARE REAL VALUED PARAMETERS WITH DIMENSIONS $(\mbox{MASS})^2$ and $(\mbox{MASS})^4$, RESPECTIVELY.}
 $H$ \lowercase{IS A DIMENSIONLESS REAL CONSTANT IN EQ(3).}  
        
E\lowercase{QUATIONS OF MOTION THAT ENSUE FROM EXTREMIZING} $S$ are given by,
$$ \psi ^{\ ; \alpha}_{\ ; \alpha}  = {1\over {2}} \bigg [ g^{\mu \nu} {{\psi_{,\mu } \psi_{,\nu}}\over{\psi}} + {B\over {A}} \psi  
-  {{H}\over{ 4 A}}\psi \  w^{\mu \nu \alpha \beta}   R^\tau_{\ \sigma \alpha \beta}  R^\sigma_{\ \tau \mu \nu} \bigg ] \eqno(4)$$ 
and,
$$R_{\mu \nu} - {1\over{2}} g_{\mu \nu} R = {{8 \pi}\over{m^2_{Pl}}} [T_{\mu \nu} + \bar{\Theta} _{\mu \nu} + C_{\mu \nu}] \eqno(5)$$
where \lowercase{THE SCALAR FIELD $\psi $ IS RELATED TO $\phi$} simply as $\psi \equiv {\phi \over {\sqrt{-g}}} $ while,     
$$\bar{\Theta} _{\mu \nu} = 2 A \bigg [{{\psi_{,\mu } \psi_{,\nu}}\over{\psi}} - {{g_{\mu \nu}}\over {2}} \bigg ( g^{\alpha \beta} {{\psi_{, \alpha } \psi_{,\beta}}\over{\psi}} + {B\over {A}} \psi \bigg )  \bigg]\eqno(6)$$
and $C^{\mu \nu}$ \lowercase{IS THE STANDARD} Cotton tensor [6], 
$$C^{\mu \nu} \equiv 2{{H}\over{\sqrt{-g}}} \bigg [ (\ln \psi)_{; \alpha ; \beta} \bigg ( *R^{\beta \mu \alpha \nu} + *R^{\beta \nu \alpha \mu} \bigg ) - (\ln \psi)_{; \alpha}
\bigg (\epsilon^{\alpha \mu \sigma \tau} R^\nu _{\ \sigma ; \tau} + \epsilon^{\alpha \nu \sigma \tau} R^\mu _{\ \sigma ; \tau} \bigg ) \bigg ]  \eqno(7)$$
with,        
$$*R R \equiv {1\over{2}} \epsilon^{\mu \nu \alpha \beta}  R^\tau_{\ \sigma \alpha \beta}  R^\sigma_{\ \tau \mu \nu} \ \ \ \ \ \ \ \ \ \ \ \ \ \ \ *R^{\tau \rho \mu \nu} \equiv {1\over{2}} \epsilon^{\mu \nu \alpha \beta} R^{\tau \rho}_ {\ \ \alpha \beta} \ \ .$$

    
     
 F\lowercase{OR A} smooth FRW universe, the Chern-Simons \lowercase{EXTENSION HAS NO EFFECT SINCE THE} Cotton \lowercase{TENSOR VANISHES} identically.   
 But \lowercase{INHOMOGEINITY AND ANISOTROPY RESULTING FROM GRAVITATIONAL COLLAPSE OF COLD DARK MATTER WOULD RESULT IN
  A NON-VANISHING} $*R R$ on smaller length scales, \lowercase{WHICH WOULD ACT AS A SOURCE OF THE DYNAMICAL FOUR-FORM DUE TO EQ.(4).}
 T\lowercase{HIS
  CAN LEAD TO A LATE TIME ACCELERATED EXPANSION OF THE UNIVERSE IF}  $A$ and $B$ have suitable values [3].
If $B < 0$, \lowercase{ONE CAN INTERPRET $\pm \sqrt {\psi} $ TO BE A QUINTESSENCE FIELD WITH A MASS} $\mu = \sqrt {\frac {-B} {4 A}}$.
 A Chern-Simons \lowercase{EXTENSION TO ELECTRODYNAMICS FOLLOWS NATURALLY FROM THE DYNAMICAL FOUR-FORM.} I\lowercase{TS EFFECT ON MAGNETOHYDRODYNAMICS EQUATIONS 
IN THE CONTEXT OF} Cowling's \lowercase{THEOREM AND PULSARS HAS BEEN EXAMINED [7].}
        
\section{Propagation of Gravitational Waves}
    
 In this section, \lowercase{WE STUDY THE EFFECT OF THE FOUR-FORM ON GRAVITATIONAL WAVES} (GWs) \lowercase{BY ASSUMING THAT, EXCEPT FOR A WEAK } GW and $\psi $,
\lowercase{THERE IS NO OTHER MATTER PRESENT.} Adopting a TT-gauge for the plane
GW travelling along the x-axis,
    $$ h_{\oplus}\equiv h_{22} (t,x)= - h_{33} (t,x) \ \ \ \ \ h_{\otimes}\equiv h_{23} (t,x)=  h_{32} (t,x)$$
   we express $S$ of eq.(2), using $g_{\mu \nu} = \eta_{\mu \nu} + h_{\mu \nu}$ and 
\lowercase{RETAINING ONLY UPTO QUADRATIC TERMS IN $ h_{\oplus} $ AND $ h_{\otimes}$ FOR THE GRAVITATIONAL PART OF THE ACTION [6,8],}
$$S \approx - {{m^2_{Pl}}\over{8 \pi}} \int {[h_{\oplus}  (\ddot{h}_{\oplus} - h^{\prime \prime}_{\oplus} ) +  h_{\otimes}  (\ddot{h}_{\otimes} - h^{\prime \prime}_{\otimes} )]d^4x} +
S_{GCS} + S_\phi \eqno(8)$$
         with,
$$S_{GCS}= - H \int{\eta^{\tau \lambda} \bigg ( \frac {\partial^2 \ln \psi}{\partial x \partial x^\tau} ( h_{\oplus} h_{\otimes, 0 \lambda} - 
 h_{\otimes} h_{\oplus, 0 \lambda}) + \frac {\partial^2 \ln \psi}{\partial t \partial x^\tau} (h_{\otimes} h_{\oplus, 1 \lambda} -  h_{\oplus} h_{\otimes, 1\lambda})
 \bigg )d^4x}  $$ 
$$\ \ \ \ \ \ \ \ \ - H\int {\frac {\partial \ln \psi} {\partial x} \bigg (h_{\oplus}  (\ddot{h}_{\otimes, 0} - h^{\prime \prime}_{\otimes,0} ) - h_{\otimes}  (\ddot{h}_{\oplus, 0} - h^{\prime \prime}_{\oplus,0})\bigg )d^4x} $$ 
$$\ \ \ \ \ \ \ \ \ + H \int{ \frac {\partial \ln \psi} {\partial t} \bigg (h_{\oplus}  (\ddot{h}_{\otimes, 1} - h^{\prime \prime}_{\otimes,1} ) - h_{\otimes}  (\ddot{h}_{\oplus, 1} - h^{\prime \prime}_{\oplus,1})\bigg )d^4x}$$
         and,
$$S_\phi = \int {\bigg [ \frac {A} {\psi} \eta^{\mu \nu} \psi_{ ,\mu} \psi_{ ,\nu} + B \psi - \frac {A} {\psi} \bigg ( h_\oplus (\psi_{,2}\psi_{,2} - \psi_{,3}\psi_{,3})
 + 2 h_\otimes \psi_{,2}\psi_{,3}\bigg ) \bigg] d^4x}$$
    
    E\lowercase{QUATIONS OF MOTION THAT FOLLOW FROM} variation of $S$ (eq.(8)) with respect to $ h_{\oplus}$, $h_{\otimes}$ and $\psi $ are given by,
$$\ddot{h}_\oplus - h^{\prime \prime}_\oplus= -\frac{8 \pi}{m^2_{Pl}} \bigg [ H \bigg \lbrace \frac {\partial^2 \chi}{\partial t \partial x} (
\ddot{h}_{\otimes} + h^{\prime \prime}_{\otimes}) - h_{\otimes,0 1}\bigg (\frac{\partial^2 \chi} {\partial t^2} + \frac{\partial^2 \chi} {\partial x^2}\bigg ) 
+ \frac {\partial \chi}{\partial x} (\ddot{h}_{\otimes, 0} - h^{\prime \prime}_{\otimes,0})  - \frac {\partial \chi}{\partial t} (
\ddot{h}_{\otimes, 1} - h^{\prime \prime}_{\otimes,1}) \bigg \rbrace + $$
$$\ \ \ \ \ \ \ \ \ +  A \exp{\chi}\bigg \lbrace \bigg (\frac {\partial \chi}{\partial y} \bigg )^2 - \bigg (\frac {\partial \chi}{\partial z} \bigg )^2 \bigg \rbrace \bigg]\eqno(9)$$
    
$$\ddot{h}_\otimes - h^{\prime \prime}_\otimes= \frac{8 \pi}{m^2_{Pl}} \bigg [ H \bigg \lbrace \frac {\partial^2 \chi}{\partial t \partial x} (
\ddot{h}_{\oplus} + h^{\prime \prime}_{\oplus}) - h_{\oplus,0 1}\bigg (\frac{\partial^2 \chi} {\partial t^2} + \frac{\partial^2 \chi} {\partial x^2}\bigg ) 
+ \frac {\partial \chi}{\partial x} (\ddot{h}_{\oplus, 0} - h^{\prime \prime}_{\oplus,0})  - \frac {\partial \chi}{\partial t} (
\ddot{h}_{\oplus, 1} - h^{\prime \prime}_{\oplus,1}) \bigg \rbrace + $$
$$\ \ \ \ \ \ \ \ \ \  + 2A \exp{\chi}\frac {\partial \chi}{\partial y} \frac {\partial \chi}{\partial z}  \bigg] \eqno(10)$$
 \lowercase{AND}
$$\partial^\mu \partial_\mu \chi + \frac {1} {2} \eta^{\mu \nu} \frac {\partial \chi}{\partial x^\mu} \frac {\partial \chi}{\partial x^\nu} - \frac {B} {4 A} = h_\oplus \bigg [\frac {\partial^2 \chi}{\partial y^2} - \frac {\partial^2 \chi}{\partial z^2} + \frac {1} {2} \bigg (\frac {\partial \chi}{\partial y}\bigg )^2
-  \frac {1} {2} \bigg (\frac {\partial \chi}{\partial z} \bigg )^2 \bigg ] +2 h_\otimes \bigg [\frac {\partial^2 \chi}{\partial y \partial z} + \frac {1} {2} \frac {\partial \chi}{\partial y} \frac {\partial \chi}{\partial z} \bigg ] \eqno(11)$$
where $\chi \equiv \ln \psi $.
T\lowercase{HE DERIVATIVES WITH RESPECT TO TIME $t$ AND X-COORDINATE ARE DENOTED BY DOT AND PRIME, RESPECTIVELY.}
        
 A\lowercase{ SIMPLIFICATION OCCURS IF WE ASSUME $\psi $ TO DEPEND ONLY
   ON THE X-COORDINATE AND TIME, LIKE THE} GW amplitude.   
  W\lowercase{ITH $\chi = \chi (t,x)$, THE LAST TERMS IN EQS.(9) AND (10) DROP OUT.} I\lowercase{NTRODUCING  THE COMPLEX CIRCULARLY POLARIZED} GW \lowercase{AMPLITUDES [8],}
    $$h_R = \frac {1} {\sqrt {2}} (h_\oplus + i h_\otimes)\ \ \ \ \ \ h_L = \frac {1} {\sqrt {2}} (h_\oplus - i h_\otimes)$$
\lowercase{ONE FINDS FROM EQS.(9) AND (10) THAT,}
$$\ddot{h}_R - h^{\prime \prime}_R= -\frac{8 \pi H i}{m^2_{Pl}} \bigg [\dot{\chi} \frac {\partial} {\partial x} (\ddot{h}_R - h^{\prime \prime}_R)
 - \chi^\prime  \frac {\partial} {\partial t} (\ddot{h}_R - h^{\prime \prime}_R) + \dot {h}^\prime_R (\ddot{\chi} + \chi^{\prime \prime}) - 
 \dot{\chi}^\prime (\ddot{h}_R + h^{\prime \prime}_R) \bigg ]\eqno(12)$$ 
    $$\ddot{h}_L - h^{\prime \prime}_L= \frac{8 \pi H i}{m^2_{Pl}} \bigg [\dot{\chi} \frac {\partial} {\partial x} (\ddot{h}_L - h^{\prime \prime}_L)
 - \chi^\prime  \frac {\partial} {\partial t} (\ddot{h}_L - h^{\prime \prime}_L) + \dot {h}^\prime_L (\ddot{\chi} + \chi^{\prime \prime}) - 
 \dot{\chi}^\prime (\ddot{h}_L + h^{\prime \prime}_L) \bigg ]\eqno(13)$$ 
     
  S\lowercase{INCE $\chi = \chi (t,x)$, THE EQUATION OF MOTION FOR THE FOUR-FORM GIVEN BY EQ.(11) SIMPLIFIES TO,}
$$\ddot{\chi} - \chi^{\prime \prime} + \frac {1} {2} (\dot{\chi}^2 - \chi^{\prime\ 2}) + 2 \mu^2=0 \eqno(14)$$
   where $\mu \equiv \sqrt{- \frac {B} {4 A} }$ acts as the  mass of the four-form field [3].
 S\lowercase{EEKING EXACT SOLUTIONS, INVOLVING MONOCHROMATIC AND CIRCULARLY POLARIZED} GWs, we substitute,
 $$h_R(t,x)=h \exp(i(\omega t- kx))$$
\lowercase{IN EQS.(12), LEADING TO A RELATION,}
$$ \omega^2 - k^2= -\beta [k (\omega^2 - k^2) \dot{\chi} + \omega (\omega^2 - k^2) \chi^\prime - i \omega k (\ddot{\chi} + \chi^{\prime \prime})
 - i (\omega^2 + k^2) \dot{\chi}^\prime ]\eqno(15)$$
 where $\beta \equiv \frac{8 \pi H }{m^2_{Pl}}$.

I\lowercase{F $w=\pm k$, EQ.(15) IMPLIES,}
$$ 2\dot{\chi}^\prime \pm  (\ddot{\chi} + \chi^{\prime \prime})=0 \ \ .\eqno(16)$$
The only self-consistent solution of eqs.(14) and (16) is,
$$\chi(t,x)= (a_0 - \frac {\mu^2} {a_0}) t + (a_0 + \frac {\mu^2} {a_0}) x \eqno(17)$$
\lowercase{WHERE $a_0$ IS AN INTEGRATION CONSTANT.} B\lowercase{UT THE ABOVE SOLUTION IS UNPHYSICAL, AS IT  IMPLIES
 AN EXPONENTIALLY GROWING $\psi $.}
 On the other hand, when $w^2 > k^2$,
  $$k \dot{\chi} + \omega \chi^\prime + \frac {1} {\beta} =0 \Rightarrow \chi^{\prime \prime} = \frac {k^2} {w^2}\  \ddot{\chi}\eqno(18)$$  
\lowercase{AND THEREFORE,} $\chi = \chi (x - \frac {\omega} {k} t)$, leading to an exact solution for $a_1 \geq \lambda$,
    $$\chi= \ln {\bigg (\sqrt{\frac {a_1} {\lambda}}\cos^2 \bigg (\pm \sqrt{\frac{\lambda} {2}} (t - \frac {k} {\omega}  x) + a_2 \bigg ) \bigg )}\eqno(19)$$
where $\lambda \equiv \frac{2 \omega^2 \mu^2} {\omega^2 - k^2}$, \lowercase{$a_1$ and $a_2$ ARE INTEGRATION CONSTANTS.}
Hence, $\psi =\exp (\chi) \propto \cos^2 \bigg (\pm \sqrt{\frac{\lambda} {2}} (t - \frac {k} {\omega}  x) + a_2 \bigg )$, with $| \omega | > | k |$, is an
acceptable solution.
I\lowercase{NDEED, IT IS POSSIBLE TO HAVE PLANE AND MONOCHROMATIC} GWs \lowercase{WITH PHASE VELOCITY EXCEEDING THE SPEED OF LIGHT.}
However, \lowercase{EQ.(14) BEING A NONLINEAR DIFFERENTIAL EQUATION, OBTAINING EXACT SOLUTIONS CORRESPONDING TO} GWs \lowercase{WITH SUPERPOSED WAVE MODES IS NON-TRIVIAL.}
        
\section{Summary}
  
G\lowercase{RAVITATIONAL WAVES PROPAGATING
   IN THE PRESENCE OF A DYNAMICAL FOUR-FORM CAN BE DECOUPLED USING CIRCULARLY POLARIZED WAVEFORMS.} I\lowercase{N THIS PAPER, WE HAVE OBTAINED EXACT, SELF-CONSISTENT
 SOLUTIONS, AND HAVE SHOWN THAT THE PHASE VELOCITY
 OF A MONOCHROMATIC GRAVITATIONAL WAVE CAN GENERICALLY EXCEED THE SPEED OF LIGHT.}
\section{References}
[1] Miao Li, Xiao-Dong Li, Shuang Wang and Yi Wang 2011 {\it Commun.Theor.Phys.} {\bf 56} 525 

[2] Das Gupta P 2012 {\it Resonance} {\bf 17} 254 

[3] Das Gupta P Dark energy and Chern-Simons like gravity from a dynamical four-form {\it Preprint} gr-qc/09051621 

[4] Guendelman E I and  Kaganovich A B 1999 {\it Phys. Rev.} D {\bf 60} 065004

[5] Gronwald F, Muench U,  Macıas A and  Hehl F W 1998 {\it Phys. Rev.} D {\bf 58} 084021

[6] Jackiw R and Pi S -Y 2003 {\it Phys. Rev.} D {\bf 68} 104012

[7] Das Gupta P 2010 On Chern-Simons corrections to magnetohydrodynamics equations {\it Radiation Effects and Defects in Solids: Incorporating 
Plasma Science and Plasma Technology} {\bf 165} 106 
({\it Preprint} gr-qc/09113967)

[8] Alexander S and Martin J 2005 {\it Phys. Rev.} D {\bf 71} 063526




 






       
       

       
\end{document}